\documentclass[conference]{IEEEtran}
\IEEEoverridecommandlockouts

\usepackage{cite}
\usepackage{amsmath,amssymb,amsfonts}
\usepackage{algorithmic}
\usepackage{graphicx}
\usepackage{textcomp}
\usepackage{xcolor}
\usepackage{setspace}
\usepackage{subcaption}
\usepackage{enumerate}
\usepackage{hyperref}
\usepackage{booktabs}
\usepackage{multirow}
\usepackage{bigstrut}
\def\BibTeX{{\rm B\kern-.05em{\sc i\kern-.025em b}\kern-.08em
    T\kern-.1667em\lower.7ex\hbox{E}\kern-.125emX}}

\begin{document}

\title{\textbf{Attention-Enhanced Short-Time Wiener Solution for Acoustic Echo Cancellation}
}

\author{\IEEEauthorblockN{Fei Zhao, Xueliang Zhang}
\IEEEauthorblockA{
\textit{College of Computer Science, Inner Mongolia University, 
Hohhot, China} \\
zhaofei@mail.imu.edu.cn, cszxl@imu.edu.cn}

}

\fontsize{9.1pt}{11.3pt}\selectfont

\maketitle

\begin{abstract}
Acoustic Echo Cancellation (AEC) is an essential speech signal processing technology that removes echoes from microphone inputs to facilitate natural-sounding full-duplex communication. 
Currently, deep learning-based AEC methods primarily focus on refining model architectures, frequently neglecting the incorporation of knowledge from traditional filter theory.
This paper presents an innovative approach to AEC by introducing an attention-enhanced short-time Wiener solution. 
Our method strategically harnesses attention mechanisms to mitigate the impact of double-talk interference, thereby optimizing the efficiency of knowledge utilization. The derivation of the short-term Wiener solution, which adapts classical Wiener solutions to finite input causality, integrates established insights from filter theory into this method.
The experimental outcomes corroborate the effectiveness of our proposed approach, surpassing other baseline models in performance and generalization.
The official code is available at \href{https://github.com/ZhaoF-i/ASTWS-AEC}{https://github.com/ZhaoF-i/ASTWS-AEC}

\end{abstract}

\begin{IEEEkeywords}
Acoustic echo cancellation, attention-enhanced, short-time Wiener solution.
\end{IEEEkeywords}

\section{Introduction}
\label{sec:intro}

Acoustic Echo Cancellation (AEC) is an important technology in speech signal processing. This technology eliminates far-end sound in full duplex communication to avoid users hearing their own voice repeatedly \cite{sondhi1967adaptive, benesty2001advances, enzner2014acoustic, hansler2005acoustic}. 
Traditional filtering techniques, exemplified by the Least Mean Squares (LMS) and Recursive Least Squares (RLS) algorithms \cite{DBLP:journals/ejasp/PaleologuCBG15,paleologu2015overview, haykin2005adaptive}, have received considerable interest as viable solutions for AEC. These methodologies are recognized for their iterative fitting process to model the echo path, which endows them with a high degree of generalizability. Nonetheless, this generality can impede their convergence performance in dynamic scenarios where the echo path is subject to change. Moreover, these techniques are challenged by their inability to effectively manage non-linear echoes, a limitation pronounced when employing low-fidelity audio equipment such as low-quality speakers and microphones.

In recent years, the rapid advancement of deep learning has promoted the emergence of numerous deep learning-based AEC methods. These approaches exhibit an enhanced capacity to address nonlinear issues, thereby mitigating some of the limitations inherent in traditional AEC techniques. Notably, Zhang et al. \cite{DBLP:conf/icassp/ZhangLZ22, DBLP:journals/taslp/ZhangLL023} proposed in-place convolution recurrent neural networks (ICRN), which utilize in-place convolution and channel-wise temporal modeling for preserving the near-end signal information. Zhang et al. \cite{zhang2022multi} proposed MTFAA, a multi-scale time-frequency processing, and streaming axial attention-based approach, exemplifying the trend towards model refinement within the field. While the prevalence of model-centric improvements has driven significant advancements in the field, there is a risk of overlooking the enduring theoretical advantages that traditional filtering techniques contribute to signal processing. These established methods offer a foundational understanding and a robust framework that complements modern approaches.

In the context of the AEC-Challenge \cite{DBLP:journals/corr/abs-2309-12553, DBLP:conf/icassp/CutlerSPPGBSA22}, many models have integrated results from traditional signal processing as part of their input strategy. While this practice has demonstrated the potential to elevate performance, the overall efficiency of utilization remains suboptimal. This is largely attributed to the significant interference encountered by traditional methods when confronted with a double-talk scenario, which in turn impinges on the model's performance.

More recently, the field of AEC has witnessed the introduction of hybrid approaches \cite{revach2021kalmannet,yang2023low,DBLP:journals/corr/abs-2301-12363}. These innovative methods amalgamate deep learning prowess with the established foundations of traditional signal processing techniques. The core premise involves employing compact neural networks to predict filter parameters, thereby striving to attain more optimal outcomes. However, the limited scale of the neural networks employed in these hybrid approaches constrains the capacity to fully exploit the potential of deep learning, thereby inherently imposing an upper limit on the attainable performance improvements.

This paper introduces an innovative AEC method that integrates short-term Wiener solutions with an attention mechanism. To address the adverse effects of double-talk scenarios, we utilize an attention mechanism for preprocessing the input of the short-term Wiener solutions. This preprocessing step enables the model to concentrate on the more salient features during single-talk instances, thereby enhancing the effective utilization of information present in clear speech segments.
The short-term Wiener solution, a variant of the traditional Wiener solution designed for finite and causal inputs, serves as a foundational component of the input signal. It incorporates traditional filter theory insights, providing a robust basis for our AEC method.
The AEC model presented in this paper represents an enhancement of the established ICCRN \cite{liu2023iccrn} framework. The experimental results conclusively demonstrate that our proposed method achieves significant enhancements in performance compared to the base models, and surpasses the efficacy of other established baseline models.

\section{PROBLEM FORMULATION}
\label{sec:FORMULATION}


\subsection{AEC system}
In the context of AEC, the system typically has access to two critical input signals: the near-end microphone signal $d(n)$ and the far-end signal $x(n)$. The formulation of the microphone signal $d(n)$ in the time domain is presented as follows:
\begin{equation}
     d(n) = y(n) + s(n)
\end{equation}
where the acoustic echo y(n) is x(n) convolving with a room impulse response (RIR) \cite{habets2006room} h(n), s(n) is the near-end speech signal.
AEC is usually applied to the time-frequency representation of the signal, we reformulate Eq.(1) into the T-F domain by applying a short-time Fourier transform (STFT) as
\begin{equation}
\label{2}
    D[t, f] = S[t, f] + \sum_k^K H[k,f]X[t-k, f]
\end{equation}
where $Y[t, f]$, $S[t, f]$ and $X[t, f]$ represent the microphone signal, near-end signal, and far-end signal at the frame $t$ and frequency $f$, respectively, and $H[k, f]$ is the echo path. Here, $K$ represents the total number of blocks.

\begin{figure}[t]
        \centering
	\includegraphics[width=1\linewidth]{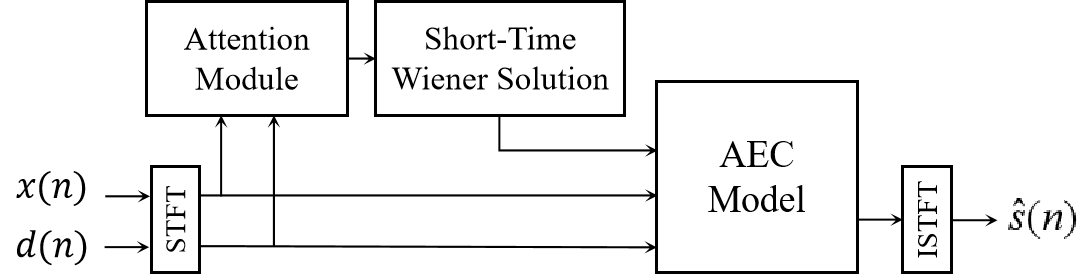}
	\caption{The schematic flowchart of the overall structure.}
	\label{fig:overview}
\end{figure}

\subsection{Wiener solution}
In an acoustically stable environment, the echo path delay, room reverberation, and volume fluctuations can effectively be characterized as a series of linear transformations applied to the far-end signal. These linear transformations are aptly captured by the mathematical abstraction of a linear discrete Finite Impulse Response (FIR) filter: 
\begin{equation}
    y'(n) = \sum_{i=0}^{N-1} x(n-i) h_i = {x_i}^\mathrm{T} h_i
\end{equation}
where $h_i$ denotes the $i$-th coefficient of the filter and $\mathrm{T}$ is a transpose operation. In the absence of any sound at the near end except for the echo signal, the signal received is effectively the echo signal itself, hence $d(n)=y(n)$. This scenario is commonly referred to as 'far-end single talk.' Under such circumstances, the error in our echo estimation, denoted as $e(n)$, can be mathematically expressed as follows:
\begin{equation}
    e(n) = y(n) - y'(n)
\end{equation}
Wiener solution aims to determine the filter coefficients $h_i$ that minimize the mean square error (MSE) between the estimated and actual echo signals. The cost function for solving the minimum MSE is as follows.
\begin{equation}
\begin{split}
\label{5}
    J_{min}(k_i) &= E\{e^2(n)\} \\
     &= E\{|y(n) - y'(n)|^2\} \\
     &= E\{|y(n) - {x_i}^\mathrm{T} h_i|^2\}
\end{split}
\end{equation}
By computing the derivative of \autoref{5}, we derive the following expression:
\begin{equation}
\label{6}
    2E\{{x_i}^T x_i\}h_i - 2E\{{x_i}^T y(n) \} = R_i h_i - r_i = 0 
\end{equation}
\begin{equation}
\label{7}
    h_i = {R_i}^{-1}  r_i
\end{equation}
where $R_i$ represents the auto-correlation matrix of the far-end signal $x_i$, with ${R_i}^{-1}$ representing its inverse. The $r_i$ denotes the cross-correlation matrix between the far-end $x_i$ and the echo signal $y(n)$. The solution $h_i$, which is directly  computed, is recognized as the Wiener solution.

\section{PROPOSED METHOD}
\label{sec:METHOD}

\subsection{Overall structure}
As depicted in \autoref{fig:overview}, our proposed method encompasses three integral components: The Attention Module, which is designed to refine the input for the short-time Wiener solution by selectively emphasizing relevant features. The short-time Wiener solution, which transforms the traditional Wiener solution into a real-time version, facilitating its application in dynamic acoustic environments. The AEC Module, which integrates the output from the short-time Wiener solution with the inputs of the conventional AEC model. This synergy serves as the input for the current AEC module, enhancing its capability to predict the near-end speech with improved accuracy.


\begin{figure}[t]
        \centering
	\includegraphics[width=0.85\linewidth]{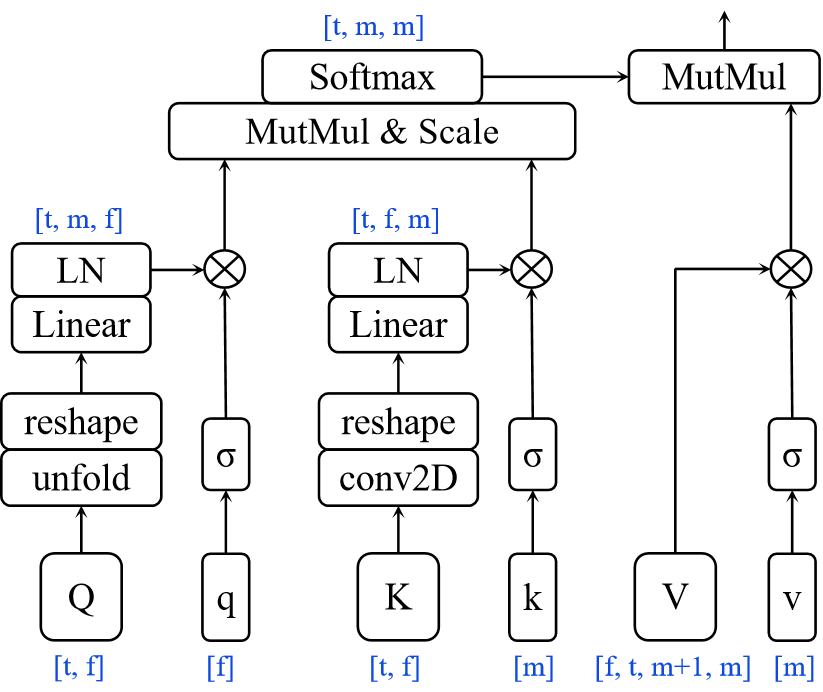}
	\caption{Diagram of the attention-enhanced Wiener solution input.}
	\label{fig:attention}
\end{figure}
\subsection{Attention-enhanced Wiener solution input}
\label{sssec:attention-enhanced}
In light of our network's operation on signals in the time-frequency (T-F) domain, we have adapted the short-time Wiener solution to this domain. Consequently, Equations (6) is reformulated to express their equivalents in the T-F domain as follows:
\begin{equation}
\label{8}
    2E\{X^T X\}H - 2E\{X^T Y\} = 0 
\end{equation}
Where $X$, $Y$, and $H$ are the far-end signal, microphone signal, and filter coefficients in the T-F domain, respectively. The far-end signal matrix $X$ is of shape $[f, t, m]$, where $t$ corresponds to the time dimension, encompassing the current frame along with the preceding $m - 1$ frames, $f$ signifies the frequency dimension, and $m$ represents the number of filter blocks. The microphone signal matrix $Y$ has the shape $[f, t]$, and the filter coefficient matrix H is shaped as $[f, n]$.

In traditional filtering techniques, the estimation of the filter during double-talk scenarios often leads to convergence issues, where the filter fails to accurately adapt to the correct parameters. This can result in echo leakage or degradation of the near-end signal. To address this challenge, we employ an attention mechanism to augment the input of our short-term Wiener solution. By leveraging the inherent properties of the attention mechanism, we aim to accentuate the single-talk intervals and mitigate the adverse effects of double-talk on the filtering process.

As depicted in \autoref{fig:attention}, the attention module is designed with three learnable vectors, $q$, $k$, and $v$, which interact with their respective inputs $Q$, $K$ and $V$. Within this framework, $Q$ signifies the far-end signal in the T-F domain, while $K$ denotes the microphone signal in the same domain. $V$ is constructed as a combination of $XTX$ and $XTY$, as articulated in \autoref{8}. The refinement of $Q$, $K$ and $V$ is facilitated by the learnable vectors $q$, $k$, and $v$ through gating mechanisms. To compute the short-time Wiener solution and capture the information from the preceding $m - 1$ frames, we perform an unfolding operation on the remote signal within the T-F domain. This operation is essential for transforming the signal into a format that can be effectively utilized by the Wiener solution. Subsequently, to ensure that the vector $K$ conforms to the requirements for attention mechanism calculations, we employ point-wise convolution to expand its channel dimension to $m$. This expansion aligns $K$ with the necessary structure for the subsequent attention-based processing.
The mathematical expression encapsulating this process is presented as follows:
\begin{align}
    Q_1 &= LN(Linear(unfold(Q))) \odot \sigma (q) \\
    K_1 &= LN(Linear(conv2D(K))) \odot \sigma (k) \\
    V_1 &= V \odot \sigma (v) \\
    W &= \frac{Q_1 {K_1}^T}{\sqrt{m}}  \\
    A &= Softmax(W)V_1
\end{align}
where $\odot$ symbolizes element-wise multiplication. The function $\sigma$ denotes the sigmoid activation function, which is utilized to introduce non-linearity and squash the output between 0 and 1. The acronym 'LN' refers to Layer Normalization. Finally, the matrix $A$ embodies the output of the attention mechanism, serving as an enhanced representation of the original signals $X^TX$ and $X^TY$. These enhanced versions are denoted as $X^TX_1$ and $X^TY_1$, respectively, indicating that the attention mechanism has selectively emphasized certain features of the input signals.

\subsection{Short-time Wiener solution}
\label{sssec:STWS}
While the Wiener solution derived from \autoref{8} promises the highest degree of accuracy, its computation entails inverting a large matrix, an operation that is excessively computationally demanding. Moreover, any modification in the echo path mandates a recalculation of the Wiener solution. These factors render the Wiener solution impractical for real-time audio interaction scenarios.

To address these limitations, we introduce a short-time Wiener solution approach. This approach confines the input length for the Wiener solution to a fixed and present-only scope, thereby preventing any leakage of future information. This constraint substantially reduces the computational burden associated with matrix inversions and bolsters the solution's adaptability to changes in the echo path. 

We incorporate the matrices $X^TX_1$ and $X^TY_1$, as derived in \autoref{sssec:attention-enhanced}, into \autoref{8} to compute the attention-enhanced short-term Wiener solution, denoted by the matrix $H^W$. Subsequently, we substitute $H^W$ into \autoref{2} to derive the following result:
\begin{equation}
   S^W[t, f] =  D[t, f] - \sum_{k=0}^{m-1} H^W[k,f]X[t-k, f]
\end{equation}
Where $S^W$ is the near-end speech predicted by $H^W$. Subsequently, $S^W$, in conjunction with the far-end and microphone signals in the T-F domain, serves as inputs to the model. Our AEC model is structured upon the ICCRN \cite{liu2023iccrn} framework and incorporates a single-layer codec for streamlined processing. 

\subsection{Loss function}
\label{sssec:Loss-function}
In this method, the corresponding loss function consists of multiple items. Stretched Scale-Invariant Signal-to-Noise Ratio (S-SISNR)\cite{DBLP:conf/interspeech/SunYZH21} is a modified version of the Scale-Invariant Signal-to-Noise Ratio (SISNR)\cite{DBLP:journals/taslp/LuoM19} loss function. S-SISNR is a time domain loss function that is obtained by doubling the period of SISNR. The simplified formula for S-SISNR is expressed as follows:
\begin{equation}
	\mathcal{L}_{\text {s-sisnr }}=10 \log_{10} cot^2(\frac{\beta}{2}) = 10\log_{10}\frac{1+cos(\beta)}{1-cos(\beta)}
\end{equation}
where $\beta$ represents the angle between two vectors' ideal near-end signal $s$ and predicted near-end signal $\hat{s}$, since it is complicated to calculate the half angle, after the derivation of the trigonometric function, it can be represented by $cos(\beta)$.

The “RI+Mag” loss criterion is adopted to recover the complex spectrum as follows:
\begin{equation}
	\mathcal{L}_{\mathrm{mag}}=\frac{1}{TF} \sum_t^T \sum_f^F ||S(t, f)|^p-|\hat{S}(t, f)|^p|^2 
\end{equation}

\begin{equation}
	\mathcal{L}_{\mathrm{RI}}=\frac{1}{TF} \sum_t^T \sum_f^F ||S(t, f)|^p e^{j \theta_{S(t, f)}}-|\hat{S}(t, f)|^p e^{j \theta_{\hat{S}(t, f)}}|^2
\end{equation}
where $p$ is a spectral compression factor (set to 0.5).
Operator $\theta$ calculates the phase of a complex number. Then the total loss function is as follows:
\begin{equation}
	\mathcal{L}_{total}=\mathcal{L}_{\mathrm{RI}}+\mathcal{L}_{\text{mag }}+\mathcal{L}_{\text{s-sisnr}}
\end{equation}

\section{EXPERIMENTAL SETUP}
\label{sec:EXPERIMENT}

\subsection{Datasets}
The near-end and far-end signals employed in our experiments are sourced from the ICASSP 2023 AEC challenge's synthetic datasets, representing near-end and far-end scenarios, respectively \cite{DBLP:journals/corr/abs-2309-12553}. The RIR is generated using the Image method \cite{allen1979image}. We simulate different rooms of size $l \times w \times h $ $m^3$ for training mixtures, where $l$ ranges from 3 to 8m, $w$ from 3 to 7, and $h$ from 3 to 5, each incremented by 0.5m. To emphasize the effectiveness of RIR prompts, the microphone-loudspeaker (M-L) distance is fixed from [0.2, 0.3, 0.4, 0.5, 0.8]m. The reverberation time (T60) is randomly selected from [0.1, 0.2, 0.3, 0.4, 0.5, 0.6]s to generate RIRs in each room. Then echo speech $v(n)$ is mixed with near-end speech $s(n)$ at signal-to-echo ratio (SER) randomly chosen from [-10, 10]dB with step 1dB. The nonlinearity setting adheres to the configuration established in \cite{DBLP:conf/icassp/ZhangLZ22}. We have generated a dataset comprising 200,000 training samples, 10,000 validation samples, and 1,000 test samples via a randomized selection process, with non-linear settings accounting for $90\%$. In addition, we also use the ICASSP 2023 AEC Challenge blind test set to evaluate the performance and generalization of our method. All speech samples have a duration of 5 seconds and the sampling rate is 16kHz. 


\begin{table*}
\belowrulesep=-0.4pt
\aboverulesep=0pt
\caption{The performance, computational complexity, and parameter count of the proposed method are compared with different baseline models in a synthetic test set. Among them, DT represents double-talk, and ST\_FE represents far-end single-talk. Using ERLE \cite{enzner2014acoustic}, PESQ \cite{DBLP:conf/icassp/RixBHH01} and SDR \cite{vincent2006performance} as the metrics.}
\centering
\scalebox{1.05}{
\fontsize{10}{14}\selectfont
\begin{tabular}{c|ccccccc|cc}
\toprule
Test scenarios       & \multicolumn{6}{c}{DT} &      \multicolumn{1}{c|}{ST\_FE}  &  \multirow{3}[5]{*}{Macs(G)} & \multirow{3}[5]{*}{Param(M)}   \\
\cmidrule(lr){2-7}\cmidrule(lr){8-8} 
SER   & \multicolumn{2}{c}{-10}  & \multicolumn{2}{c}{0}  & \multicolumn{2}{c}{10}     & \multicolumn{1}{c|}{--}    \\
\cmidrule(lr){2-3}\cmidrule(lr){4-5}\cmidrule(lr){6-7}\cmidrule(lr){8-8}
Model\textbackslash{}Metric  & PESQ   & SDR  & PESQ   & SDR & PESQ   & SDR & ERLE    \\
\midrule
mix                                      & 1.33               & -10        & 1.93        & 0              & 2.56        & 10         & --      & --      & --       \\
ICRN                                                 & 2.69               & 3.06       & 3.37        & 3.63           & 3.83        & 2.91       & 36.07   & 1.78    & 0.213    \\
MTFAA                                                & 2.66               & 9.83       & 3.30         & 15.37          & 3.78        & 22.3       & 50.4    & 5.419   & 2.149    \\
\midrule
ICCRN                                               & 2.76               & 10.64      & 3.43        & 12.93          & 3.88        & 13.60       & 45.02   & 0.844   & 0.12     \\
STWS                                          & \textbf{2.89}              & 11.68      & 3.54        & 14.14          & 3.96        & 14.91      & 48.72   & 0.85    & 0.12     \\
\textbf{ASTWS}                                         & \textbf{2.89}               & \textbf{15.77}      & \textbf{3.56}       & \textbf{23.46}          & \textbf{4.00}       & \textbf{29.63}      & \textbf{56.41}   & 0.963   & 0.148        \\

\bottomrule
\end{tabular}
}
\label{synthetic}
\end{table*}

\begin{table}
\belowrulesep=-0.4pt
\aboverulesep=0pt
\caption{Comparison of the proposed method with the baseline model on the ICASSP2023 blind test set.}
\centering
\scalebox{1.05}{
\fontsize{10}{13}\selectfont
\begin{tabular}{c|cc}
\toprule
~~~Test scenarios~~~       & \multicolumn{1}{c}{~~~DT~~~} &      \multicolumn{1}{c}{~~~ST\_FE~~~} \\
\cmidrule(lr){2-2}\cmidrule(lr){3-3}
Model\textbackslash{}Metric        & \multicolumn{2}{c}{MOS\_ECHO} \\
\midrule
ICRN        & 3.718     & 2.86   \\
MTFAA                                                & 4.009     & 3.11   \\
ICCRN                                               & 2.866     & 2.193  \\
STWS                                          & 2.917     & 2.214  \\
\textbf{ASTWS}     & \textbf{4.275}     & \textbf{3.782}  \\
\bottomrule
\end{tabular}
}
\label{blind}
\end{table}

\subsection{Training details}

Our proposed method has been evaluated against a set of baseline models to ascertain its comparative performance. The baseline models include the ICRN \cite{DBLP:conf/icassp/ZhangLZ22}, the MTFAA \cite{zhang2022multi}, and the ICCRN \cite{liu2023iccrn} with a single-layer codec. To ensure a fair comparison and to accommodate the 16 kHz speech data, we have modified the MTFAA by removing its band decomposition and band merging modules. All the aforementioned baseline models, including our proposed method, adhere to the same training strategy, utilize the same training loss function, and are trained on an identical dataset.

For $m$ in \autoref{sssec:STWS}, we set it as 20.
For the STFT used to calculate the “RI+Mag” loss ($\mathcal{L}_{\mathrm{mag}}$, $\mathcal{L}_{\mathrm{RI}}$), the window size was set to 20 ms with an offset of 5 ms using a hamming window. For the STFT used within ICCRN, we utilized the default configuration provided with that model.
The model is optimized by Adam algorithm \cite{DBLP:journals/corr/KingmaB14}.
The initial learning rate is set to 0.001. If the validation loss does not decrease for two consecutive epochs, the learning rate is reduced by half. Training is stopped when the verification loss does not decrease for 10 consecutive epochs.

\section{EXPERIMENTAL RESULTS AND DISCUSSION}
\label{sec:RESULTS}

Our proposed method integrates an \textbf{A}ttention-enhanced \textbf{S}hort-\textbf{T}ime \textbf{W}iener \textbf{S}olution. To denote this integration, we refer to our method as \textbf{ASTWS}. Additionally, we conducted an ablation study to isolate the effects of the short-term Wiener solution by excluding the attention module, which we designate as STWS.
Thus, the set of experiments including ICRN, MTFAA, ICCRN, and ASTWS serves to compare the performance of our proposed method with that of the baseline models. Concurrently, the ablation experiments comprising ICCRN, STWS, and ASTWS are designed to elucidate the individual contributions of each module within our proposed method.

In \autoref{synthetic}, we present a comparative analysis of our proposed method against other baseline models on a synthetic test set. The results indicate that our method consistently achieves optimal performance across various test scenarios, SER, and evaluation metrics, with a particularly notable improvement in the SDR score. This underscores the effectiveness of our approach.

In the ablation study, it is observed that the ICCRN model experiences a significant enhancement in performance upon integration with the short-term Wiener solution, denoted as STWS. When employing the full proposed method, denoted as ASTWS, we observe a modest improvement in the PESQ index and a substantial enhancement in the SDR. These results suggest that the distinct modules within our method synergistically contribute to the overall performance improvement.
Furthermore, our proposed method demonstrates a significantly lower computational complexity and parameter count compared to the ICRN and MTFAA models. It achieves a greater performance improvement at a reduced computational expense relative to the ICCRN, highlighting its efficiency and effectiveness.

We extend our comparative analysis to the ICASSP 2023 blind test set, as detailed in \autoref{blind}. Given that our training process did not account for the addition of noise, we focus our comparison on the MOS\_ECHO index within the AECMOS evaluation metrics. The results reveal that the MTFAA model, attributed to its larger model size, exhibits superior generalization, with its MOS\_ECHO index surpassing those of other baseline models. Nonetheless, our proposed method demonstrates an even higher MOS\_ECHO index than MTFAA and far exceeds other baseline models. This outcome not only proves the effectiveness of our method but also robustly evidences its enhanced generalization capabilities.

\section{CONCLUSIONS}
\label{sec:CONCLUSIONS}

In this paper, we introduce an innovative approach to acoustic echo cancellation by proposing an attention-enhanced short-time Wiener solution. Our method leverages the attention mechanism to refine the input of the short-time Wiener solution, effectively focusing on single-speaking segments while mitigating the adverse effects of double-speaking scenarios. Subsequently, we employ a real-time, limited-input Wiener solution, termed the short-time Wiener solution, to integrate traditional filter signals into the deep learning-based acoustic echo cancellation model. The experimental results demonstrate that our proposed method not only significantly outperforms the established baseline models but also exhibits commendable generalization performance across various testing conditions.

\textbf{Acknowledgments}: This research was partly supported by the China National Nature Science Foundation (No. 61876214).

\newpage

\bibliographystyle{IEEEbib}
\bibliography{refs}

\begin{thebibliography}{10}

\bibitem{sondhi1967adaptive}
MM~Sondhi,
\newblock ``An adaptive echo canceller,''
\newblock {\em Bell System technical journal}, vol. 46, no. 3, pp. 497--511, 1967.

\bibitem{benesty2001advances}
Jacob Benesty, Tomas G{\"a}nsler, Dennis~R Morgan, M~Mohan Sondhi, Steven~L Gay, et~al.,
\newblock ``Advances in network and acoustic echo cancellation,''
\newblock 2001.

\bibitem{enzner2014acoustic}
Gerald Enzner, Herbert Buchner, Alexis Favrot, and Fabian Kuech,
\newblock ``Acoustic echo control,''
\newblock in {\em Academic press library in signal processing}, vol.~4, pp. 807--877. Elsevier, 2014.

\bibitem{hansler2005acoustic}
Eberhard H{\"a}nsler and Gerhard Schmidt,
\newblock {\em Acoustic echo and noise control: a practical approach},
\newblock John Wiley \& Sons, 2005.

\bibitem{DBLP:journals/ejasp/PaleologuCBG15}
Constantin Paleologu, Silviu Ciochina, Jacob Benesty, and Steven~L. Grant,
\newblock ``An overview on optimized {NLMS} algorithms for acoustic echo cancellation,''
\newblock {\em {EURASIP} J. Adv. Signal Process.}, vol. 2015, pp. 97, 2015.

\bibitem{paleologu2015overview}
Constantin Paleologu, Silviu Ciochin{\u{a}}, Jacob Benesty, and Steven~L Grant,
\newblock ``An overview on optimized nlms algorithms for acoustic echo cancellation,''
\newblock {\em EURASIP Journal on Advances in Signal Processing}, vol. 2015, pp. 1--19, 2015.

\bibitem{haykin2005adaptive}
Simon~S Haykin,
\newblock {\em Adaptive filter theory},
\newblock Pearson Education India, 2005.

\bibitem{DBLP:conf/icassp/ZhangLZ22}
Chenggang Zhang, Jinjiang Liu, and Xueliang Zhang,
\newblock ``A complex spectral mapping with inplace convolution recurrent neural networks for acoustic echo cancellation,''
\newblock in {\em {IEEE} International Conference on Acoustics, Speech and Signal Processing, {ICASSP} 2022, Virtual and Singapore, 23-27 May 2022}. 2022, pp. 751--755, {IEEE}.

\bibitem{DBLP:journals/taslp/ZhangLL023}
Chenggang Zhang, Jinjiang Liu, Hao Li, and Xueliang Zhang,
\newblock ``Neural multi-channel and multi-microphone acoustic echo cancellation,''
\newblock {\em {IEEE} {ACM} Trans. Audio Speech Lang. Process.}, vol. 31, pp. 2181--2192, 2023.

\bibitem{zhang2022multi}
Guochang Zhang, Libiao Yu, Chunliang Wang, and Jianqiang Wei,
\newblock ``Multi-scale temporal frequency convolutional network with axial attention for speech enhancement,''
\newblock in {\em ICASSP 2022-2022 IEEE International Conference on Acoustics, Speech and Signal Processing (ICASSP)}. IEEE, 2022, pp. 9122--9126.

\bibitem{DBLP:journals/corr/abs-2309-12553}
Ross Cutler, Ando Saabas, Tanel P{\"{a}}rnamaa, Marju Purin, Evgenii Indenbom, Nicolae{-}Catalin Ristea, Jegor Guzvin, Hannes Gamper, Sebastian Braun, and Robert Aichner,
\newblock ``{ICASSP} 2023 acoustic echo cancellation challenge,''
\newblock {\em CoRR}, vol. abs/2309.12553, 2023.

\bibitem{DBLP:conf/icassp/CutlerSPPGBSA22}
Ross Cutler, Ando Saabas, Tanel P{\"{a}}rnamaa, Marju Purin, Hannes Gamper, Sebastian Braun, Karsten S{\o}rensen, and Robert Aichner,
\newblock ``{ICASSP} 2022 acoustic echo cancellation challenge,''
\newblock in {\em {IEEE} International Conference on Acoustics, Speech and Signal Processing, {ICASSP} 2022, Virtual and Singapore, 23-27 May 2022}. 2022, pp. 9107--9111, {IEEE}.

\bibitem{revach2021kalmannet}
Guy Revach, Nir Shlezinger, Ruud~JG Van~Sloun, and Yonina~C Eldar,
\newblock ``Kalmannet: Data-driven kalman filtering,''
\newblock in {\em ICASSP 2021-2021 IEEE International Conference on Acoustics, Speech and Signal Processing (ICASSP)}. IEEE, 2021, pp. 3905--3909.

\bibitem{yang2023low}
Dong Yang, Fei Jiang, Wei Wu, Xuefei Fang, and Muyong Cao,
\newblock ``Low-complexity acoustic echo cancellation with neural kalman filtering,''
\newblock in {\em ICASSP 2023-2023 IEEE International Conference on Acoustics, Speech and Signal Processing (ICASSP)}. IEEE, 2023, pp. 1--5.

\bibitem{DBLP:journals/corr/abs-2301-12363}
Yixuan Zhang, Meng Yu, Hao Zhang, Dong Yu, and DeLiang Wang,
\newblock ``Kalmannet: {A} learnable kalman filter for acoustic echo cancellation,''
\newblock {\em CoRR}, vol. abs/2301.12363, 2023.

\bibitem{liu2023iccrn}
Jinjiang Liu and Xueliang Zhang,
\newblock ``Iccrn: Inplace cepstral convolutional recurrent neural network for monaural speech enhancement,''
\newblock in {\em ICASSP 2023-2023 IEEE International Conference on Acoustics, Speech and Signal Processing (ICASSP)}. IEEE, 2023, pp. 1--5.

\bibitem{habets2006room}
Emanuel~AP Habets,
\newblock ``Room impulse response generator,''
\newblock {\em Technische Universiteit Eindhoven, Tech. Rep}, vol. 2, no. 2.4, pp. 1, 2006.

\bibitem{DBLP:conf/interspeech/SunYZH21}
Yuhang Sun, Linju Yang, Huifeng Zhu, and Jie Hao,
\newblock ``Funnel deep complex u-net for phase-aware speech enhancement,''
\newblock in {\em Interspeech 2021, 22nd Annual Conference of the International Speech Communication Association, Brno, Czechia, 30 August - 3 September 2021}, Hynek Hermansky, Honza Cernock{\'{y}}, Luk{\'{a}}s Burget, Lori Lamel, Odette Scharenborg, and Petr Motl{\'{\i}}cek, Eds. 2021, pp. 161--165, {ISCA}.

\bibitem{DBLP:journals/taslp/LuoM19}
Yi~Luo and Nima Mesgarani,
\newblock ``Conv-tasnet: Surpassing ideal time-frequency magnitude masking for speech separation,''
\newblock {\em {IEEE} {ACM} Trans. Audio Speech Lang. Process.}, vol. 27, no. 8, pp. 1256--1266, 2019.

\bibitem{allen1979image}
Jont~B Allen and David~A Berkley,
\newblock ``Image method for efficiently simulating small-room acoustics,''
\newblock {\em The Journal of the Acoustical Society of America}, vol. 65, no. 4, pp. 943--950, 1979.

\bibitem{DBLP:conf/icassp/RixBHH01}
Antony~W. Rix, John~G. Beerends, Michael~P. Hollier, and Andries~P. Hekstra,
\newblock ``Perceptual evaluation of speech quality (pesq)-a new method for speech quality assessment of telephone networks and codecs,''
\newblock in {\em {IEEE} International Conference on Acoustics, Speech, and Signal Processing, {ICASSP} 2001, 7-11 May, 2001, Salt Palace Convention Center, Salt Lake City, Utah, USA, Proceedings}. 2001, pp. 749--752, {IEEE}.

\bibitem{vincent2006performance}
Emmanuel Vincent, R{\'e}mi Gribonval, and C{\'e}dric F{\'e}votte,
\newblock ``Performance measurement in blind audio source separation,''
\newblock {\em IEEE transactions on audio, speech, and language processing}, vol. 14, no. 4, pp. 1462--1469, 2006.

\bibitem{DBLP:journals/corr/KingmaB14}
Diederik~P. Kingma and Jimmy Ba,
\newblock ``Adam: {A} method for stochastic optimization,''
\newblock in {\em 3rd International Conference on Learning Representations, {ICLR} 2015, San Diego, CA, USA, May 7-9, 2015, Conference Track Proceedings}, Yoshua Bengio and Yann LeCun, Eds., 2015.

\end{thebibliography}

\end{document}